\documentstyle[emulateapj]{article}

\def\Mpc{h^{-1} \, \mbox{Mpc}}

\def\etal{{\it et al. }}
\def\spose#1{\hbox to 0pt{#1\hss}}
\def\simlt{\mathrel{\spose{\lower 3pt\hbox{$\mathchar''218$}}
     \raise 2.0pt\hbox{$\mathchar''13C$}}}
\def\simgt{\mathrel{\spose{\lower 3pt\hbox{$\mathchar''218$}}
     \raise 2.0pt\hbox{$\mathchar''13E$}}}
\def\beq{\begin{equation}}
\def\eeq{\end{equation}}
\def\bce{\begin{center}}
\def\ece{\end{center}}
\def\bea{\begin{eqnarray}}
\def\eea{\end{eqnarray}}
\def\ben{\begin{enumerate}}
\def\een{\end{enumerate}}

\def\nn{\nonumber}

\def\brr{\begin{array}}
\def\err{\end{array}}

\newcommand{\rhobar}{\overline{\rho}}

\newcommand{\xibar}{\overline{\xi}}

\def\Or{{\cal O}}

\def\calH{{\cal H}}

\font\twelveBF=cmmib10 scaled 1000

\newcommand{\x}{\hbox{\twelveBF x}}

\newcommand{\vv}{\hbox{\twelveBF v}}

\newcommand{\rb}{\hbox{\twelveBF r}}
\newcommand{\vu}{\hbox{\twelveBF u}}
\newcommand{\lexp}{\mathop{\bigl\langle}}
\newcommand{\rexp}{\mathop{\bigr\rangle}}
\newcommand{\rexpc}{\mathop{\bigr\rangle_c}{}}

\def\ga{\mathrel{\mathpalette\fun >}}
\def\fun#1#2{\lower3.6pt\vbox{\baselineskip0pt\lineskip.9pt
\ialign{$\mathsurround=0pt#1\hfill##\hfil$\crcr#2\crcr\sim\crcr}}}

\begin{document} 

\title{Non-Linear Gravitational growth of large scale structures  \\
inside and outside standard Cosmology}

\author{E. Gazta\~naga$^{a,b~~}$\footnotemark[1]\addtocounter{footnote}{1},
J.A. Lobo$^{c~~}$\footnotemark[2]\addtocounter{footnote}{1}}
\affil{$^{a}$ INAOE, Astrofisica, Tonantzintla, 
Apdo Postal 216 y 51, 7200, Puebla, Mexico}
\affil{$^{b}$ Institut d'Estudis Espacials de Catalunya,  
IEEC/CSIC, Gran Capit\'an 2-4, 08034 Barcelona, Spain}
\affil{$^{c}$ Departament de Fisica Fonamental, Universitat de
Barcelona, Diagonal 647, 08028 Barcelona, Spain}



\begin{abstract}

We reconsider the problem of gravitational structure formation inside
and outside General Relativity (GR), both in the weakly and strongly
non-linear regime. We show how these regimes can be explored
observationally through clustering of high order cumulants and through
the epoch of formation, abundance and clustering of collapse
structures, using Press-Schechter formalism and its extensions. We
address the question of how different are these predictions when using
a non-standard theory of Gravity. We study examples of cosmologies
that do not necessarily obey Einstein's field equations: scalar-tensor
theories (STT), such as Brans-Dicke (BD), parametrized with $\omega$,
a non-standard parametrisation of the Hubble law, $H^2=
a^{-3(1+\epsilon)}$, or a non-standard cosmic equation of state
$p=\gamma\rho$, where $\gamma$ can be chosen irrespective of the
cosmological parameters ($\Omega_M$ and $\Omega_\Lambda$). We present
some preliminary bounds on $\gamma$, $\omega$ and $\epsilon$ from
observations of the skewness and kurtosis in the APM Galaxy
Survey. This test is independent of the overall normalization of rms
fluctuations. We also show how abundances and formation times change
under these assumptions. Upcoming data on non-linear growth will place
strong constraints on such variations from the standard paradigm.

\end{abstract}


\section{Introduction}

\footnotetext[1]{gazta@inaoep.mx, gaztanaga@ieec.fcr.es}
\footnotetext[2]{lobo@ffn.ub.es}

In Cosmology the standard picture of gravitational growth, and also many
aspects of fundamental physics, are extrapolated many orders of magnitude,
from the scales and times where our current theory of gravity (General
Relativity, GR) has been experimentally tested, into the distant universe.
In particular, current limits on the (parametrized) Post Newtonian formalism
mostly restrict to our very local Universe (see Will 1993). It is important
to evaluate  how much our predictions and cosmological picture depend on the
underlying hypothesis (see Peebles 1999 for insightful comments on the state
of this subject). The other side of this argument is that cosmology can be
used to test fundamental physics, such as our theory of gravity.

One aspect of GR that could be questioned or tested without modifying the
basic structure or symmetry of the theory are Einstein's field equations,
relating the energy content ($T_{\mu\nu}$) to the curvature ($R_{\mu\nu}$).
One such modification, which will be considered here, is scalar-tensor
theories (STT), such as Brans-Dicke (BD) theory. A more generic, but also
more vague, way of testing the importance of Einstein's field equations is
to model independently the geometry and the matter content, thus allowing
for the possibility of other relations between them. Some simple aspects
of this idea will be illustrated here by studying structure formation
in a flat, matter dominated universe but with a more general growth law for
the Hubble rate ---see section \ref{sec32} below. Similarly, we will also
consider results for a generic equation of state: $p=\gamma \rho$, where
$\gamma$ can be chosen independently of the cosmological parameters
($\Omega_M$, $\Omega_k$ and $\Omega_\Lambda$).

Our aim in this paper is to explore certain variations of 
the standard model to see how
they affect structure formation. The idea is to find a way to parameterize
variations from GR that might produce differences large enough to be
observable. The variations considered could have other observable
consequences (eg in the local universe or in the radiation dominated regime)
which might rule them out as a viable new theory. But even if this were the
case, we still would have learn something about how structure formation 
depends on the underlying theory of Gravity or the assumptions about the
equation of state. This aspect of the theory has hardly been explored and it
therefore represents an important  step forward in analyzing alternatives to
the current paradigm, eg non-baryonic matter (see Peebles 1999), and could
also help to set limits on variations of GR or the equation of state at high
red-shifts.

Here we consider two main regimes for structure formation in
non-standard gravity/cosmology: weakly non-linear and strongly
non-linear large scale clustering. We study the shear-free or
spherical collapse (SC) model, which corresponds to the spherically
symmetric (or local) dynamics (see below). This approximation works
very well at least in two different contexts, that will be explored
here.
 
The first one is the growth of the smoothed 1-point cumulants of the
probability distribution for large scale density fluctuations: the SC
model turns out to reproduce exactly the leading order perturbation
theory predictions (Bernardeau 1992), and turns out to be an excellent
approximation for the exact dynamics as compared to N-body simulations
both with Gaussian (Fosalba \& Gazta\~naga 1998a, 1998b) and non-Gaussian
initial conditions (Gazta\~naga \& Fosalba 1998). The measured 1-point
cumulants in galaxy catalogues have been compared with these predictions
(eg Bouchet \etal 1993, Gazta\~naga 1992,1994,  Gazta\~naga \& Frieman 1994,
Baugh, Gazta\~naga \& Efstathiou 1995, Gazta\~naga 1995, 
Baugh \& Gazta\~naga 1996, Colombi etal 1997, Hui \& Gazta\~naga 1999). 

The second one is the study of the epoch of formation and abundance of
structures (such as galaxies and clusters), using the Press \& Schechter
(1974) formalism and its extensions (eg  Bond \etal 1991, Lacey \& Cole
1993, Sheth \& Lemson 1999, Scoccimarro \etal 2000). 
Given some Gaussian initial conditions, this formalism can predict
the number of structures (halos) of a given mass that will form at each
stage of the evolution. One can use the SC model to predict the value of
the critical  linear over-density, $\delta_c$, that will collapse into
virialized halos. It turns out that the analytical predictions for the
halo mass function and formation rates are remarkably accurate as compared
to N-body simulations (Lacey \& Cole 1994). One can also use this type of
modeling to predict clustering properties of halos (eg Mo \& White 1996,
Mo, Jing \& White 1997), cluster abundances
(White, Efstathiou \& Frenk 1993, Bahcall \& Fan 1998) or  weak lensing 
through mass functions
(Jain \& Van Waerbeke 2000). The observed cluster
abundances have been used as a strong discriminant for cosmological models
and also as a way to measure the amplitude of mass fluctuations, $\sigma_8$
(see White, Efstathiou \& Frenk 1993, Bahcall \& Fan 1998).

In summary, we propose to address a very specific question here: how
different are the above non-linear predictions when using a
non-standard cosmology and non-standard theory of Gravity? To answer
this question we will consider two non-standard variations:
scalar-tensor models and some examples of a cosmology that do not obey
Einstein's field equations. The paper is organized as follows: In \S2
we give a summary on how non-linear structure formation relates to the
underlying theory of Gravity (see Weinberg 1972, Peebles 1993, Ellis
1999 and references therein, for a review on the relation between
gravitational theory and cosmology).  This section covers old ground
with some detail as an introduction to later sections and for the
reader that is not familiar with this subject or notation.  We also
present the more general case of an ideal (relativistic) fluid. As far
as we know, some of the non-linear results presented here are new.  In
\S3 we show how these predictions change in the two examples of
non-standard gravity. Observational consequences are explored in
\S4. In \S5 we present a discussion and the conclusions.

\section{Gravitational Growth inside GR}

The self-gravity of an over-dense region work against the expansion
of the universe so that this region will expand at a slower rate that the
background. This increases the density contrast so that eventually the
region collapses. The details of this collapse depends on the initial
density profile. Here we will focus in the spherically symmetric case.
We will revise non-linear structure growth in the context of the
fluid limit and the shear-free approximation. These turns out to be very
good approximation for the applications that will be considered later
(leading order and strongly non-linear statistics).

We start with the Raychaudhuri's equation, which is  valid for an
arbitrary Ricci tensor $R_{\mu\nu}$. We use Einstein's field equations
and the continuity equation to turn Raychaudhuri's equation into a
second order differential equation for the density contrast. We
first present the matter dominated (non-relativistic) case, with
solutions for the linear and non-linear regimes. Later,
in~\S\ref{sec:eqofstate}, we assess the more generic case of an ideal
(relativistic) fluid and its corresponding solution.

\subsection{Einstein's and Raychaudhuri's Equations}

We start recalling that the metric tensor $g_{\mu\nu}$ defines 
the line element of space-time:

\beq
ds^2 = ~g_{\mu\nu} ~dx^\mu ~dx^\nu
\eeq
which in the homogeneous and isotropic model of the cosmological
principle can be written as (see eg Weinberg 1972):

\beq
ds^2 = dt^2 - a^2(t)~\left[ {dr^2\over{1+k r^2}} + r^2 
\left( d\theta^2 + \sin^2\theta d\phi^2\right)~\right]
\label{frw}
\eeq

As usual we will work in comoving coordinates $\x$ related to
physical coordinates by $\rb_p=a(t)\,\x$, where $a(t)=(1+z)^{-1}$
is the cosmic scale factor, and $z$ the corresponding red-shift
($a_0\equiv 1$). Thus all geometrical aspects of this universal
line element are determined up to the function $a(t)$ and the
arbitrary constant $k$, which defines the usual open, Einstein-deSitter
and closed universes. The function $a(t)$ can be found for each energy
content by solving the corresponding equations of motion, eg the
gravitational field equations.

In this section we consider Einstein's equations:

\beq
R_{\mu\nu} + \Lambda g_{\mu\nu} =
-8\pi G\,\left(T_{\mu\nu}-\frac 12\, g_{\mu\nu}T\right)
\label{8}
\eeq
where $T\equiv g^{\mu\nu}\,T_{\mu\nu}$ is the trace of the energy-momentum
tensor; we have included a cosmological constant term to keep the equations
general at this stage. For an ideal fluid, we have:

\beq
T_{\mu\nu} = p\, g_{\mu\nu} + (p+\rho)\,u_\mu u_\nu
\label{9}
\eeq

We can now use the  field equations and the above
energy-momentum to find the scale factor $a(t)$ in
the metric:

\bea
\frac{3\ddot a}a & = & -4\pi G\rho\,\left(1+\frac{3p}\rho\right)
+ \Lambda       \label{ne1}     \\
H^2 \equiv \frac{\dot a^2}{a^2} & = & \frac{8\pi G\rho}3 +
\frac{k}{a^2} + \frac\Lambda 3  \label{ne2}\ ,\qquad \dot{}\equiv\frac d{dt}
\eea

In the fluid approximation, deviations from the mean background $\rhobar$
are characterized by fluctuations in the density and velocity fields. The
continuity equation for a non-relativistic fluid is (Peebles 1993):

\beq
{{\partial } \over {\partial \tau}}\,\delta (\x,\tau) + \nabla \cdot 
\{\left[{1+ \delta (\x,\tau)} \right] \vv(\x,\tau) \} = 0
\label{eq:mass}
\eeq
where $\delta(\x,\tau) \equiv \rho(\x,\tau)/\rhobar-1$ is the local
{\em density contrast}, $\vv(\x,\tau)$ the {\em peculiar velocity}
(see Eq.[\ref{vp}] below), and $\tau$ the {\em conformal time} defined by

\beq
d\tau = \frac{dt}{a(t)} \Leftrightarrow
\frac d{dt}=\frac 1a\,\frac d{d\tau}
\eeq

The continuity equation [\ref{eq:mass}] can also be written

\beq
\frac{d\delta}{d\tau} + (1+\delta)\,\theta = 0 ,\qquad
\theta\equiv\nabla\cdot\vv      \label{5}
\eeq

In order to find an equation of motion for the density contrast alone
we shall resort to the Raychaudhuri equation (see eg Wald 1984)

\beq
\frac{d\Theta}{ds} + \frac 13\,\Theta^2 = -\sigma_{ij}\sigma^{ij} +
\omega_{ij}\omega^{ij} + R_{\mu\nu}\,u^\mu u^\nu
\label{6}
\eeq
where $\Theta\equiv\nabla_\mu u^\mu$, $\sigma_{ij}$ is the {\em shear}
tensor, $\omega_{ij}$ the {\em vorticity} tensor, and $R_{\mu\nu}$ the
Ricci tensor, and $s$ the proper time parameter; $u^\mu$ is the fluid's
4-velocity, $u^0=1$, and

\beq
\vu = \dot a(t)\,\x + \vv(\x,t) \label{vp}
\eeq

It is important to stress that Raychaudhuri's equation, Eq. [\ref{6}],
is {\em purely geometric}: it describes the evolution in proper time
of the dilatation coefficient $\Theta$ of a bundle of nearby geodesics.
There is no physics in this equation until a relationship between
$R_{\mu\nu}$ and the matter contents of the universe is specified by
means of a set of field equations. This makes it very useful for our
purposes in this paper, as we shall later make reference to
a different set of field equations.

If Einstein's field equations, Eq. [\ref{8}] and [\ref{9}], are assumed
then it is readily verified that

\beq
R_{\mu\nu}\,u^\mu u^\nu = -4\pi G\rho\,\left(1+\frac{3p}\rho\right)
+ \Lambda       \label{10}
\eeq

\subsection{Shear free and matter domination}

In a matter dominated regime ($p=0$), $\rho\sim a^{-3}$. Equation
[\ref{ne2}] for the Hubble rate $H$, can be rewritten using  the notation:
$\Omega_M \equiv 8\pi G \rho_0/(3H_0^2)$, which 
is the ratio of the current matter density to the critical density,
$\Omega_k=k/H_0^2$ gives the global curvature,
and $\Omega_\Lambda=\Lambda/(3H_0^2)$ where
$\Lambda$ is the cosmological constant, so that  
$\Omega_M+\Omega_k+\Omega_\Lambda=1$:

\begin{equation}
H^2(z) =
H^2_0 ~\left[ ~\Omega_M (1+z)^{3} +
\Omega_k (1+z)^2 + \Omega_\Lambda ~\right]
\label{H}
\end{equation}

We can now replace equation [\ref{10}] into equation [\ref{6}]. 
In a matter-dominated regime, and for a shear free, 
non-rotating cosmic fluid we obtain:

\beq
\frac{d\Theta}{dt} + \frac 13\,\Theta^2 = -4\pi G\rho + \Lambda
\label{11}
\eeq

On making use of equation [\ref{vp}] we can split $\Theta$ as

\beq
\Theta\equiv\nabla_\mu u^\mu = \frac{3\dot a}a +
\frac\theta a           \label{12}
\eeq
so that, taking into consideration the field equations for the expansion
factor $a(t)$ (Eqs. [\ref{ne1}] and [\ref{ne2}]), equation [\ref{11}]
can be recast in the form

\beq
\frac{d\theta}{d\tau} + {\cal H}(\tau)\,\theta +
\frac 13\,\theta^2 = -4\pi Ga^2\,\rhobar\delta
\label{13}
\eeq
where ${\cal H}(\tau)\equiv d(\ln a)/d\tau$. We can now eliminate
$\theta$ between eqs.\ [\ref{5}] and [\ref{13}] to find the following
second order differential equation for the density contrast:

\bea
&& \frac{d^2\delta}{d\tau^2} + {\cal H}(\tau)\,\frac{d\delta}{d\tau} -
\frac 32\,{\cal H}^2(\tau)\,\Omega_M(\tau)\,\delta \nn \\ &=&
\frac 43\,(1+\delta)^{-1}\,\left(\frac{d\delta}{d\tau}\right)^2 +
\frac 32\,{\cal H}^2(\tau)\,\Omega_M(\tau)\,\delta^2
\label{14}
\eea
where we have shifted to the rhs all non-linear terms, and used the
notation

\beq
\Omega_M(\tau) =
\frac{\Omega_M}{\Omega_M + a\,\Omega_k + a^3\,\Omega_\Lambda}
\eeq

Equation [\ref{14}] reproduces the equation of the {\em spherical collapse}
model (SC). In other words, {\em the SC approximation is the exact dynamics
when shear is neglected} (see Fosalba \& Gazta\~naga 1998a). As one would
expect, this yields a {\em local} evolution, in the sense that the evolved
field at a point is just given by a local (non-linear) transformation of the
initial field at the same point, with independence of the surroundings. This
SC solution yields the exact perturbation theory predictions for the
cumulants at tree-level (leading order with Gaussian initial conditions) and
it also is an excellent approximation for next to leading orders, see below.
As mentioned in the introduction, one can also  use the SC model to predict
the value of the critical  linear overdensity, $\delta_c$, that will collapse
into virialized halos.

\begin{figure*}
\centerline{\epsfxsize=8truecm
\epsfbox{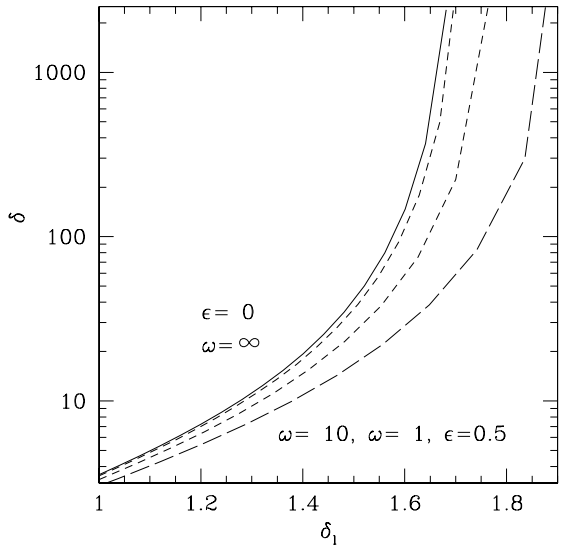}}
\figcaption[junk]{The non-linear density contrast, 
$\delta$, as a function of the
linear one $\delta_l$ in the spherical collapse. The continuous line shows
the GR prediction ($\omega=\infty$, $\epsilon=0$, $\gamma=0$), the
short-dashed lines correspond to the BD model with $\omega=10$ and
$\omega=1$ (from left to right). The long-dashed line shows the case with
a non-standard Hubble rate $H^2= a^{-3(1+\epsilon)}$ for $\epsilon=0.5$.
\label{deldell}}
\end{figure*}

\subsection{Linear growth}
\label{sec:linear}

We next do a perturbative expansion for $\delta$. The first contribution
is the linear theory solution. For this, equation [\ref{14}] clearly
simplifies to

\beq
\frac{d^2\delta_l}{d\tau^2} + \calH (\tau)\,\frac{d\delta_l}{d\tau}
-{3\over{2}} ~{\cal H}^2(\tau)\,\Omega_M(\tau) ~\delta_l = 0
\label{linear}
\eeq    
where $\delta_l$ stands for the ``linear'' solution. Because 
the coefficients of the above equation are time dependent only,
the spatial and temporal part factorise:

\beq
\delta_l(\x,\tau) = \delta_0(\x)\,D(\tau)
\label{deltal}
\eeq
where  $D$ is usually called the {\em linear growth factor}. Thus initial
fluctuations, no matter of what size, are amplified by the same factor, and
the statistical properties of the initial field are just linearly scaled.
For example, the $N$-point correlation functions are:

\beq
\xi_N(r_1,..,r_N,t) = D^N \xi_N(r_1,..,r_N,0)
\label{xiN}
\eeq

To find the solution to equation [\ref{linear}] it is expedient to change
the time variable to $\eta=\ln(a)$, so that

\beq
{d\over{d\eta}}= {1\over{{\cal H}(\tau)}}{d\over{d\tau}} =
{1\over H}{d\over{dt}}
\label{eta}
\eeq

We then have

\beq
{d^2 D \over{d^2\eta}} + \left(2 + {\dot H\over{H^2}}\right){dD\over{d\eta}}
- {3\over{2}}\;\Omega_M(\eta) ~D = 0 
\label{lineareta}
\eeq
where we can write

\bea
{\dot H\over{H^2}} &=& - {3\over{2}} \left( {{\Omega_M + 2/3 ~e^\eta~\Omega_k}
\over{\Omega_M+ e^\eta~ \Omega_k + e^{3\eta}~\Omega_\Lambda}} \right)
\\[1 em]
\Omega_M(\eta) &=&
{\Omega_M\over{\Omega_M+ e^\eta~ \Omega_k + e^{3\eta}~ \Omega_\Lambda}}
\label{oeta}
\eea
where $\Omega_M$, $\Omega_k$ and $\Omega_\Lambda$ are just constants
(the current value at $a=1$). 

In the Einstein-deSitter universe ($\Omega_k=\Omega_\Lambda=0$) we have that
$\Omega_M(\eta)=1$ and $\dot H/H^2=-3/2$, so the differential equation becomes

\beq
{d^2D\over{d^2\eta}} + {1\over{2}} ~ {d D \over{d\eta}} - {3\over{2}}~D = 0
\eeq
whose solutions

\beq
D = C_1~e^\eta  + C_2 e^{-3/2\eta} = C_1~ a + C_2 a^{-3/2}
\eeq
reproduce the usual linear growth $D \sim a$ and the decaying solutions
$D\sim a^{-3/2}$.

\subsection{Non-linear growth}

\begin{figure*}
\centerline{\epsfxsize=8truecm
\epsfbox{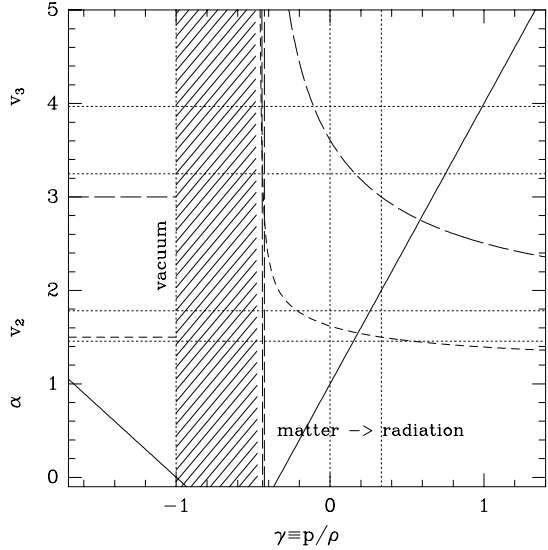}}
\figcaption[junk]{The linear growth
index $\alpha_1$ (continuous line) 
and non-linear coefficients $\nu_2$ (short-dashed)
and $\nu_3$ (long-dashed),
as a function of $\gamma \equiv p/\rho$.
Vertical dotted lines correspond to the vacuum,
matter and radiation dominated cases $\gamma=-1,0,1/3$.
The horizontal dotted lines bracketted the $\nu_2$
and  $\nu_3$ regions within
$10\%$ error of the matter dominated ($\gamma=0$) case.
\label{ggamman}}
\end{figure*}

The exact (non-perturbative) solution for the SC Eq.[\ref{14}] for the
density contrast in an Einstein-deSitter universe admits a well known
parametric representation:

\bea
\delta(\varphi) &=& {9 \over 2}{(\varphi-\sin\varphi)^2 \over
(1-\cos\varphi)^3} - 1  \nn \\  \delta_l (\varphi) &=& {3\over 5} 
\left[{3\over 4}(\varphi-\sin\varphi)\right]^{2/3}  
\label{SC1}
\eea
for $\delta_l > 0$, linear overdensity, and

\bea
\delta(\varphi) &=& {9 \over 2}{(\sinh\varphi-\varphi)^2 \over
(\cosh\varphi-1)^3} - 1  \nn \\ 
\delta_l (\varphi) &=& - {3\over 5} \left[{3\over 4}
(\sinh\varphi-\varphi)\right]^{2/3} 
\label{SC2}
\eea
for $\delta_l< 0$, linear under-density (see Peebles 1993), where the
parameter $\varphi$ is just a parametrisation of the time coordinate.
There is also a solution for the $\Omega_M\ne 1$ case (see Bernardeau
1992, Fosalba \& Gazta\~naga 1998b). The continuous line in
Figure~\ref{deldell} illustrates the solution to the above equation
(the other lines will be explained later). Note the singularity at
$\delta_l \simeq 1.686$, which corresponds to the gravitational collapse
(see \S\ref{sec:deltac} below).

If we are only interested in the perturbative regime
($\delta_l \rightarrow 0$), which is the relevant one for the
description of structure formation on large scales, the above
solution can be expressed directly in terms of the linear
density contrast, $\delta_l$, which plays the role of the initial
size of the spherical fluctuation in Eq.[\ref{deltal}]. This way,
the evolved density contrast in the perturbative regime is given 
by a {\em local-density} transformation of the
linear density fluctuation,

\beq
\delta  =  f(\delta_l) =  
\,\sum_{n=1}^{\infty} {\nu_n \over n!}\, {[\delta_l]^n}
\label{loclag}
\eeq

Notice that all the non-linear dynamical information in the SC model
is encoded in the $\nu_n$ coefficients.
We can now introduce the above power series expansion in Eq.[\ref{14}] 
and determine the $\nu_n$ coefficients one by one. Before we do this,
it is convenient to change again the time variable to $\eta=\ln(a)$
as we did in the linear case, Eq[\ref{lineareta}]:

\bea
&& {d^2\delta\over{d^2\eta}}+\left(2+{\dot H\over{H^2}}\right)
{d\delta\over{d\eta}} - {3\over{2}}\,\Omega_M(\eta) ~\delta \nn \\
&=&
{4\over{3}}\,{1\over 1+\delta}\,\left({d\delta\over{d\eta}}\right)^2
~+~ {3\over{2}}\;\Omega_M(\eta) ~\delta^2
\label{nonlineareta}
\eea

We can now use the expansion in Eq.[\ref{loclag}] with $\delta_l$
given by the linear growth factor $D=a=e^\eta$ and compare order
by order. For the Einstein-deSitter universe they turn out to be:

\beq
~~~~~~~~~~~~  \nu_2 = {34\over 21} ~~~~~~;~~~~~~ 
\nu_3 = {682\over 189}
\label{nusc}
\eeq
and so on (see eg Folsalba \& Gazta\~naga 1998b for other cases).
Once we have these coefficients we can get the
evolution of the non-linear variance and higher order moments
in terms of the initial conditions (see \S\ref{sec:cumulants} below).

\subsection{Equation of state $p=\gamma \rho$}
\label{sec:eqofstate}

\begin{figure*}
\centerline{\epsfxsize=8truecm
\epsfbox{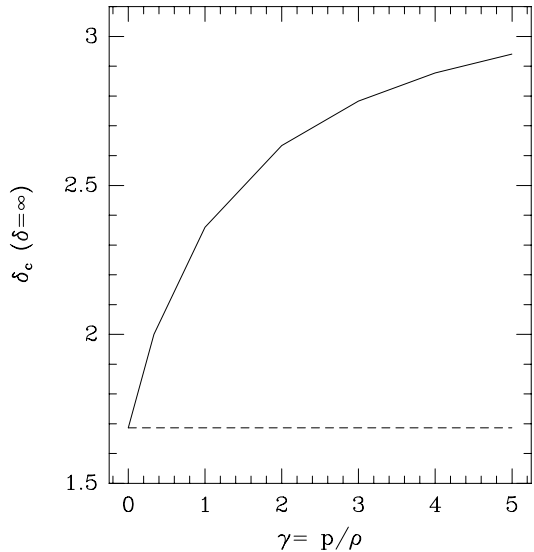}}
\figcaption{The critical value of the linear density contrast
 $\delta_c$ where $\delta = \infty$ as a function
of  $\gamma \equiv p/\rho$.
\label{deltacg}}
\end{figure*}

We will now consider a perfect fluid with equation of sate $p=\gamma \rho$.
Not all values of $\gamma$ make physical sense. Here, in the spirit of
going beyond the standard paradigm, we will ignore
these restrictions and assume that $\gamma$ can take any real constant value,
irrespective of other cosmological parameters.

The time-component of the energy conservation equations
$\nabla_{\!\nu}\,T^{\mu\nu}=0$ gives us (for $p=\gamma\rho$)
both the background density behavior
 
\beq
\rhobar\,a^{3(1+\gamma)} = \mbox{const}
\eeq
and the continuity equation for the density contrast
 
\beq
\frac{d\delta}{d\tau} + (1+\gamma)(1+\delta)\,\theta =
-\gamma\rhobar\,(\vv\cdot\nabla\delta)
\label{enerv}
\eeq
where, like before, $\tau$ is the conformal time, and
$\theta\equiv\nabla\cdot\vv$. This is the generalization of
equation~[\ref{5}] for a relativistic fluid. Note that an
additional (quadratic) term now appears in the rhs of~[\ref{enerv}].
The magnitude of this term is assessed by resorting to the 
space-components of the energy conservation equations
$\nabla_{\!\nu}\,T^{\mu\nu}=0$: these are identically
satisfied when $\gamma=0$, and they show that
$\vv\!\cdot\!\nabla\delta\propto|\vv|^2/c^2$, plus higher
order contributions. These can be safely neglected since
peculiar velocities are always very small compared to the
speed of light; in fact the approximation $|\vv|^2/c^2 \rightarrow 0$
is always made, even in the more standard case when $\gamma=0$. We
shall therefore consistently adopt the following equation
for the density contrast:
 
\beq
\frac{d\delta}{d\tau} + (1+\gamma)(1+\delta)\,\theta = 0
\label{energyc}
\eeq
 
Also, Hubble's equation, Eq. [\ref{H}], now becomes

\beq
H^2 = H_0^2\,\left[
\Omega_M\,a^{-3(1+\gamma)} + \Omega_k\,a^{-2} + \Omega_\Lambda\right]
\label{HHH}
\eeq

We can combine equation [\ref{energyc}] with the Raychaudhuri
equation for this case ---cf Eqs. [\ref{6}] and [\ref{10}]

\beq
\frac{d\Theta}{dt} + \frac 13\,\Theta^2 =
-4\pi G\rho\,\left(1+3\gamma\right) + \Lambda
\eeq
to obtain, after some algebra,

\bea
{d^2 \delta \over{d^2\eta}} + \left(2 + {\dot H\over{H^2}}\right) 
{d\delta\over{d\eta}}
- {3\over{2}} (1+\gamma)(1+3\gamma)\,\Omega(\eta)  ~\delta = && \nn \\[1 ex]
\frac{4+3\gamma}{3+3\gamma} {1\over 1+\delta}
\left({d \delta \over{d\eta}}\right)^2
~+~ {3\over{2}} (1+\gamma)(1+3\gamma)\,\Omega(\eta)~\delta^2 &&
\eea
where we have expediently redefined $\Omega(\eta)$ in Eq. [\ref{oeta}] to

\beq
\Omega_M(\eta) =
{\Omega_M\over{\Omega_M+ e^{\eta(1+3\gamma)}~ \Omega_k + e^{3\eta(1+\gamma)}~
\Omega_\Lambda}}
\eeq
and we can write
\beq
{\dot H\over{H^2}} = - {3\over{2}} \left( {{(1+\gamma)\Omega_M~e^{-3\eta\gamma} +
2/3 ~e^\eta~\Omega_k}
\over{\Omega_M~e^{-3\eta\gamma}+ e^\eta~ \Omega_k + e^{3\eta}~\Omega_\Lambda}}
\right)
\eeq
In an Einstein-deSitter universe ($\Omega_k=\Omega_\Lambda=0$),
$\Omega(\eta)=1$, and the linear regime is governed by

\beq
{d^2 D \over{d^2\eta}} + {1-3\gamma \over{2}}~ 
{dD\over{d\eta}} - {3\over{2}} (1+\gamma)(1+3\gamma)\,D = 0
\eeq
which has the usual solutions of the form $D=a^\alpha$, with

\beq
\alpha_1=1+3\gamma\ ,\qquad \alpha_2=-3(1+\gamma)/2
\eeq

Figure \ref{ggamman} shows these perturbative solutions.
The shaded region corresponds to the case where linear
evolution is suppressed, eg $\alpha<0$. In this case, as can be
seen from Eq.[\ref{alpha1g}]-[\ref{alpha2g}],
$\nu_2$ and $\nu_3$ have a very rapid variation.
The growing mode for $\gamma>-1/3$ is:

\bea
\alpha_1 &=& {1 + 3\,\gamma} \label{alpha1g} \\
\nu_2 &=& {\frac{2\,\left( 17 + 48\,\gamma + 27\,\gamma^2 \right) }
        {3\,\left( 1 + \gamma \right) \,\left( 7 + 15\,\gamma \right) }} 
\label{nu2g} \\
\nu_3 &=& \left[\, 72 + 540\,\gamma + 324\,\gamma^2 + \frac{16}{{\left( 1 + \gamma
\right) }^2} + \frac{24}{1 + \gamma} \right.  \\ \nn
 &-& \left. \frac{ \left(6\,+ 18\,\gamma \right)\left( 17 + 48\,\gamma +
27\,\gamma^2 \right)}
{\left( 1 + \gamma \right) \,\left( 7 + 15\,\gamma \right) } \right] \nn \\
&\times& \left(\,{27 + 144\,\gamma + 189\,\gamma^2}\right)^{-1}
\eea
For $\gamma<-1$ the dominant linear growth is $\alpha_2$ and
the  values of $\nu_2$ and $\nu_3$ are constant:

\bea
\alpha_2 &=&  {\frac{-3\,\left( 1 + \gamma \right) }{2}} \\
\nu_2     &=& {3\over{2}} \label{nu2g2} \\
\nu_3    &=& {3} 
\label{alpha2g}
\eea

For radiation ($\gamma=1/3$) we have that $\alpha_1=2$ which 
reproduces the well
known results (see Peebles 1993) and $\nu_2= {3\over{2}}$ and
$\nu_3=3$, which are new results as far as we know.
Note that these values are
identical to the case of negative pressure, $\gamma<-1$, the
only difference being in the linear growth, but for
$\gamma = -7/3$ all $\alpha$, $\nu_2$ and $\nu_3$
are identical to the radiation case. In the limit
of strong pressure $\gamma \rightarrow \infty$ we find:
$\nu_2 = 6/5$ and $\nu_3 = 12/7$ .
As can be seen in Figure \ref{ggamman}, and also in
the equations above, there are poles for $\nu_2$ 
at $\gamma=-1$ and $\gamma=-7/15$.

Figure \ref{deltacg} shows the corresponding variation
in $\delta_c$, defined as the value of the linear overdensity
where the corresponding non-linear value becomes infinity
(see \S \ref{sec:deltac}).

\section{Gravitational Growth outside GR}
 
\subsection{Scalar-Tensor Theories}

Here we investigate how a varying $G$ could change the above results.
We parameterize the variation of $G$ using scalar-tensor theories (STT)
of gravity such as Brans-Dicke (BD) theory or its extensions.

To make quantitative predictions we will consider cosmic evolution
in STTs, where $G$ is derived from a scalar field $\phi$ which is 
characterized by a function $\omega=\omega(\phi)$ determining the 
strength of the coupling between the scalar field and gravity. 
In the simplest BD models, $\omega$ is just a constant and $G 
\simeq \phi^{-1}$ ---see below. However if $\omega$ varies then it
can change with cosmic time, so that $\omega=\omega(z)$. The structure
of the solutions to BD equations is quite rich and depends crucially on
the coupling function $\omega(\phi)$ (see Barrow \& Parsons 1996).

Here we shall be considering the standard BD model with constant
$\omega$; the field equations are (see eg Weinberg 1972):

\bea
R_{\mu\nu} &=& -\frac{8\pi}\phi\,\left(
T_{\mu\nu}-\frac{1+\omega}{3+2\omega}\;g_{\mu\nu}T\right) -
\frac\omega{\phi^2}\,\nabla_\mu\phi\,\nabla_\nu\phi \nn \\ &-&
\frac 1\phi\,\nabla_\mu\nabla_\nu\phi  \label{bdfeq}  \\
\Box\phi & = & \frac{8\pi}{3+2\omega}\,T\ ,
\qquad (T\equiv g^{\mu\nu} T_{\mu\nu})
\eea

The Hubble rate $H$ for a homogeneous and isotropic background universe
can be easily obtained from the above; 

\begin{equation}
H^2 \equiv \left({{\dot a}\over{a}}\right)^2= {8 \pi \rho\over{3\phi}}
+{k\over{a^2}}+ {\Lambda\over{3}}+{\omega\over{6}}{{{\dot \phi}^2}
\over{\phi^2}} - H {{{\dot \phi}}\over{\phi}}
\label{31}
\end{equation}

These equations must be complemented with the equation of state for
the cosmic fluid. In a flat, matter dominated universe ($p=0$), an
exact solution to the problem can be found:

\begin{equation}
G= {4+2\omega\over{3+2\omega}}\phi^{-1} = G_0 (1+z)^{1/(1+\omega)}
\label{G(t)}
\end{equation}
and

\beq
a(t) = (t/t_0)^{(2\omega+2)/(3\omega+4)}
\label{a(t)}
\eeq 

This solution for the flat universe is recovered in a general case
in the limit $t\rightarrow\infty$, and also arises as an exact
solution of Newtonian gravity with a power law $G \propto t^{n}$
(Barrow 1996). For non-flat models, $a(t)$ is not a simple power law
and the solutions get far more complicated. To illustrate the effects
of a non-flat cosmology we will consider general solutions that
can be parametrized as Eq. [\ref{G(t)}] but which are not simple
power-laws in $a(t)$. In this case, it is easy to check that the
new Hubble law given by Eq. [\ref{31}] becomes

\begin{equation}
H^2 = H^2_0 ~\left[ ~\hat\Omega_M (1+z)^{3+1/(1+\omega)} +
\hat\Omega_k (1+z)^2 + \hat\Omega_\Lambda ~\right]
\label{H(z)}
\end{equation}
where $\hat\Omega_M$,$\hat\Omega_k$ and $\hat\Omega_\Lambda$ follow
the usual relation $\hat\Omega_M+\hat\Omega_k+\hat\Omega_\Lambda=1$,
and are related to the familiar local ratios ($z \rightarrow 0$:
$\Omega_M \equiv 8\pi G_0 \rho_0/(3H_0^2)$, $\Omega_k=k/H_0^2$
and $\Omega_\Lambda=\Lambda/(3H_0^2)$) by

\bea
\hat\Omega_M &=& \Omega_M ~
\frac{3(1+\omega)^2}{(2+\omega)(4+3\omega)}  \nn \\
\hat\Omega_\Lambda &=& \Omega_\Lambda~
\frac{6(1+\omega)^2}{(3+2\omega)~(4+3\omega)} \\ \nn 
\hat\Omega_k &=& 
\Omega_k~\frac{6(1+\omega)^2}{(3+2\omega)~(4+3\omega)}
\label{omegahat}
\eea

Thus the GR limit is recovered as $\omega\rightarrow\infty$. 

We now investigate the density fluctuations in the above theory. Like in
section II, we shall make use of the continuity equation [\ref{5}] in
combination with the Raychaudhuri equation [\ref{6}]. As mentioned above,
cf section 2.1, both of these are still valid within the context of BD
theory: it is only needed to replace the Ricci tensor in the rhs of
Eq. [\ref{6}] according to BD's field equations, Eq. [\ref{bdfeq}].
Considering again a non-rotating, shear-free cosmic fluid, we find:

\bea
&& \frac{d\Theta}{dt} + \frac 13\,\Theta^2 = \nn \\ &=&
-\frac{4+2\omega}{3+2\omega}\,
\frac{4\pi\rho}\phi\,\left(1+
\frac{1+\omega}{2+\omega}\,\frac{3p}\rho\right)-
\omega\,\frac{\dot\phi^2}{\phi^2}-\frac{\ddot\phi}\phi
\label{33.3}
\eea

We shall still make use of a gravitational ``constant'' parametrized
as in equation~[\ref{G(t)}] above; this is justified insofar as the
characteristic length for the variation of $\phi$ is typically much
greater than that of the density fluctuations in a matter dominated
universe ---see eg (Nariai 1969). In this approximation,
the above equation gives

\beq
\frac{d\theta}{d\tau} + {\cal H}(\tau)\,\theta +
\frac 13\,\theta^2 = -\frac{4+2\omega}{3+2\omega}\,
\frac{4\pi a^2\rhobar\delta}\phi
\label{33.4}
\eeq
where $\tau$ is again the {\it conformal time\/} parameter,
$d\tau=a^{-1}\,dt$, and $\theta$ is defined in equations~[\ref{5}]
and~[\ref{12}]. Remarkably, this equation is very similar to the
GR equation~[\ref{9}]: we only need to replace in it the
gravitational constant $G$ by its expression as a multiple of the
varying scalar field $\phi$ given in equation~[\ref{G(t)}].
Combining [\ref{33.4}] with the continuity equation~[\ref{5}]
we immediately find

\beq
\frac{d^2\delta}{d\tau^2} + {\cal H}(\tau)\,\frac{d\delta}{d\tau} -
\frac {4}{3\,(1+\delta)}\,\left(\frac{d\delta}{d\tau}\right)^2 =
\frac{4+2\omega}{3+2\omega}\,\frac{4\pi a^2\rho\delta}\phi
\label{33.5}
\eeq

Like in section II, we change the independent variable in [\ref{33.5}] to
$\eta=\ln a$, whereby we obtain

\bea
&& \frac{d^2\delta}{d\eta^2} + \left(2 + \frac{\dot H}{H^2}\right)\,
\frac{d\delta}{d\eta} -
\frac 43\,(1+\delta)^{-1}\,\left(\frac{d\delta}{d\eta}\right)^2 \nn \\
&=&
\frac{4+2\omega}{3+2\omega}\,\frac{4\pi a^2\rho\delta}{H^2 \phi}
\label{33.6}
\eea

Using equation [\ref{31}] to calculate $\dot H$, and assuming further
that $\hat\Omega_k=\hat\Omega_\Lambda=0$, we finally get

\bea
&& {d^2\delta\over{d^2\eta}} + {1\over{2}} ~{\omega\over{1+\omega}}
~ {d\delta\over{d\eta}}
- {1\over{2}}~\frac{(2+\omega)(4+3\omega)}{(1+\omega)^2}~\delta \nn \\
&=&   
{4\over{3}} {1\over 1+\delta}\left({d\delta\over{d\eta}}\right)^{\!2}
~+~ {1\over{2}}~\frac{(2+\omega)(4+3\omega)}{(1+\omega)^2}~\delta^2
\label{33.7}
\eea

We next examine the solutions to this equation.

\subsubsection{Linear growth}

\begin{figure*}
\centerline{\epsfxsize=8truecm
\epsfbox{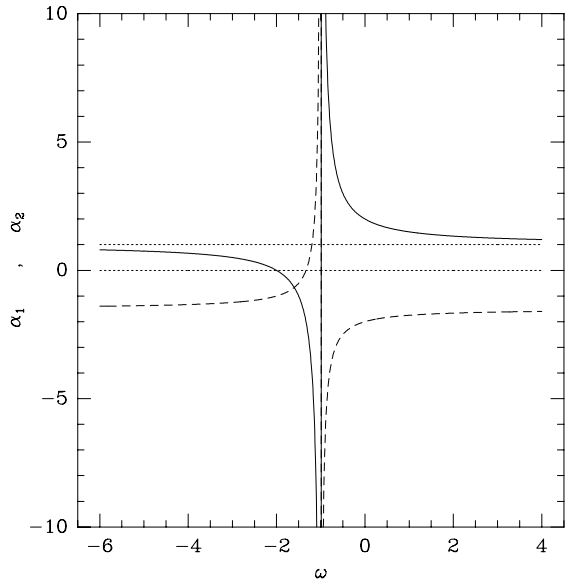}}
\figcaption[junk]{The linear growth
indices $\alpha_1$ (continuous line) and $\alpha_2$
(dashed line), 
defined by the solution $D = C_1~a^\alpha_1+C_2~a^\alpha_1$
as a function of the BD parameter $\omega$
for a time varying gravitational constant 
$G = G_0~ a^{-1/(1+\omega)}$.
\label{alphas}}
\end{figure*}

Let us call $D(\eta)$ the solution to the linearized version of equation
[\ref{33.7}], i.e.,

\beq
{d^2D\over{d^2\eta}}+{1\over{2}}~{\omega\over{1+\omega}}~{dD\over{d\eta}}
- {1\over{2}}~\frac{(2+\omega)(4+3\omega)}{(1+\omega)^2}~D = 0
\eeq

Again the solutions are given by the roots $\alpha_1$ and  $\alpha_1$
of the corresponding characteristic functions:

\beq
D = C_1~ a^{\alpha_1} + C_2 a^{\alpha_2}
\eeq
with 

\bea
\alpha_1 &=& {2+\omega \over{1+\omega}} 
~~\simeq~~ 1 + {1\over{\omega}} 
+ \Or\left({1\over{\omega^2}}\right) \label{eq:alphaw} \\
\alpha_2 &=& {-4- 3\omega \over{2+2\omega}} 
~~\simeq~~ -{3\over{2}} -{1\over 2}~ {1\over{\omega}} 
+ \Or\left({1\over{\omega^2}}\right)
\eea
which reproduces the usual linear growth $D\sim a$ and  $D\sim a^{-3/2}$
in the limit $\omega\rightarrow\infty$. Note that $\alpha_1$ corresponds
to the growing mode only for large values of $|\omega|$, but the situation
is more complicated when $\omega$ is not large.

Figure \ref{alphas} shows the values of $\alpha_1$ and $\alpha_2$ as
functions of $\omega$. The effective $G$ in BD decreases as the Universe
expands if $-1< \omega<\infty$, and the expansion factor $a(t)$ stops for
$\omega=-1$; the growing mode in this regime is controlled by $\alpha_1$,
since this is the positive root. The growing mode for $-4/3<\omega<-1$
is  $\alpha_2$, but the universe shrinks to an eventual
collapse in this regime (see Eq. [\ref{a(t)}]). Between  
$-2<\omega<-4/3$ the Universe expands again, but  there
are no growing modes, as can be seen in Figure~\ref{alphas}
(both~$\alpha_1$ and~$\alpha_2$ are negative). For $\omega<-2$ the
expansion factor grows with time and 
 $\alpha_1$ becomes the growing mode again. Notice that in this
regime of $\omega<-2$,  $\alpha_1 <1$, so that it is slower than 
for $\omega>0$. As we will show below this is compensated in part by a
stronger non-linear growth.

\subsubsection{Non-linear growth}

\begin{figure*}
\centerline{\epsfxsize=8truecm
\epsfbox{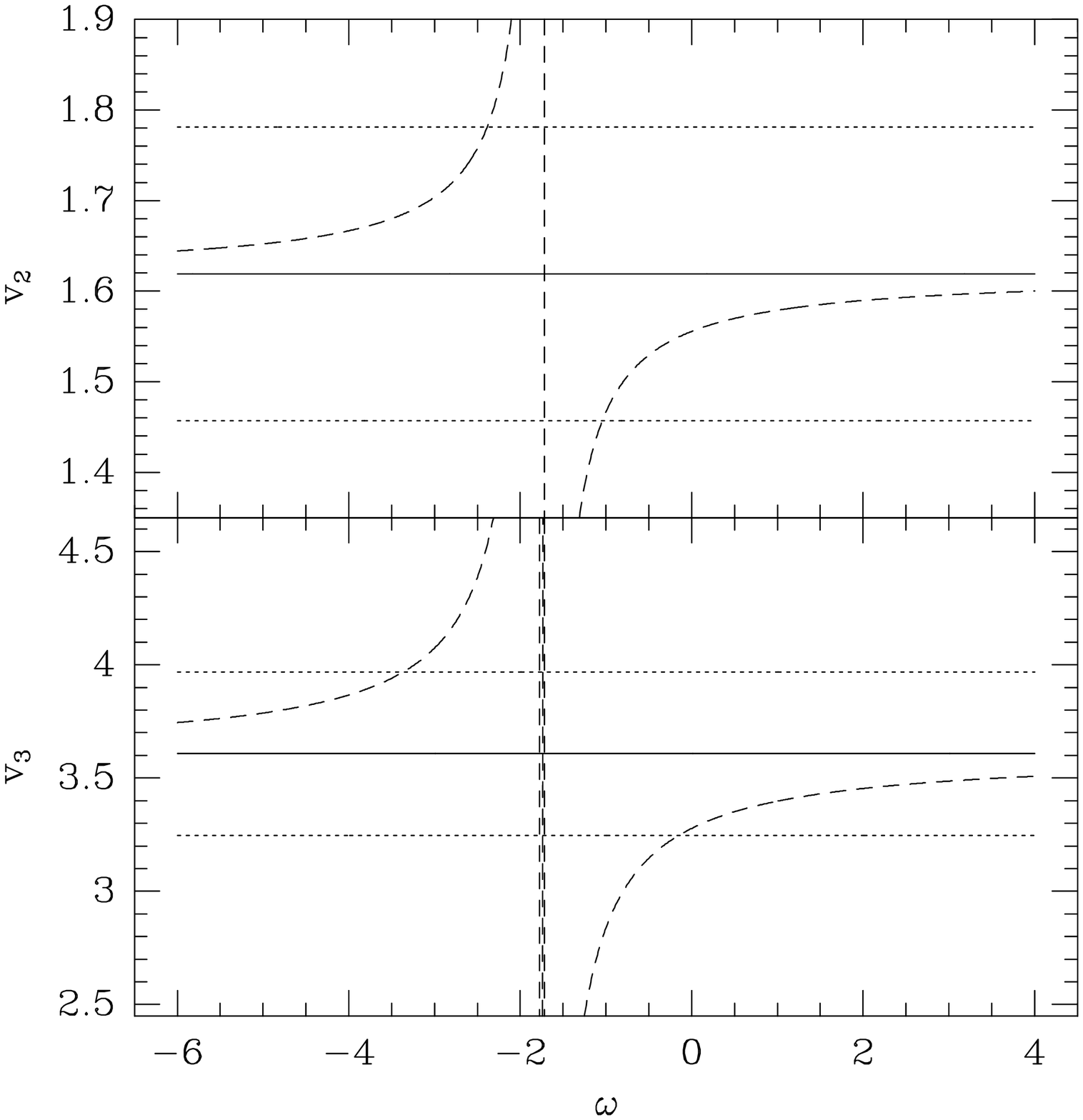}\epsfxsize=8truecm
\epsfbox{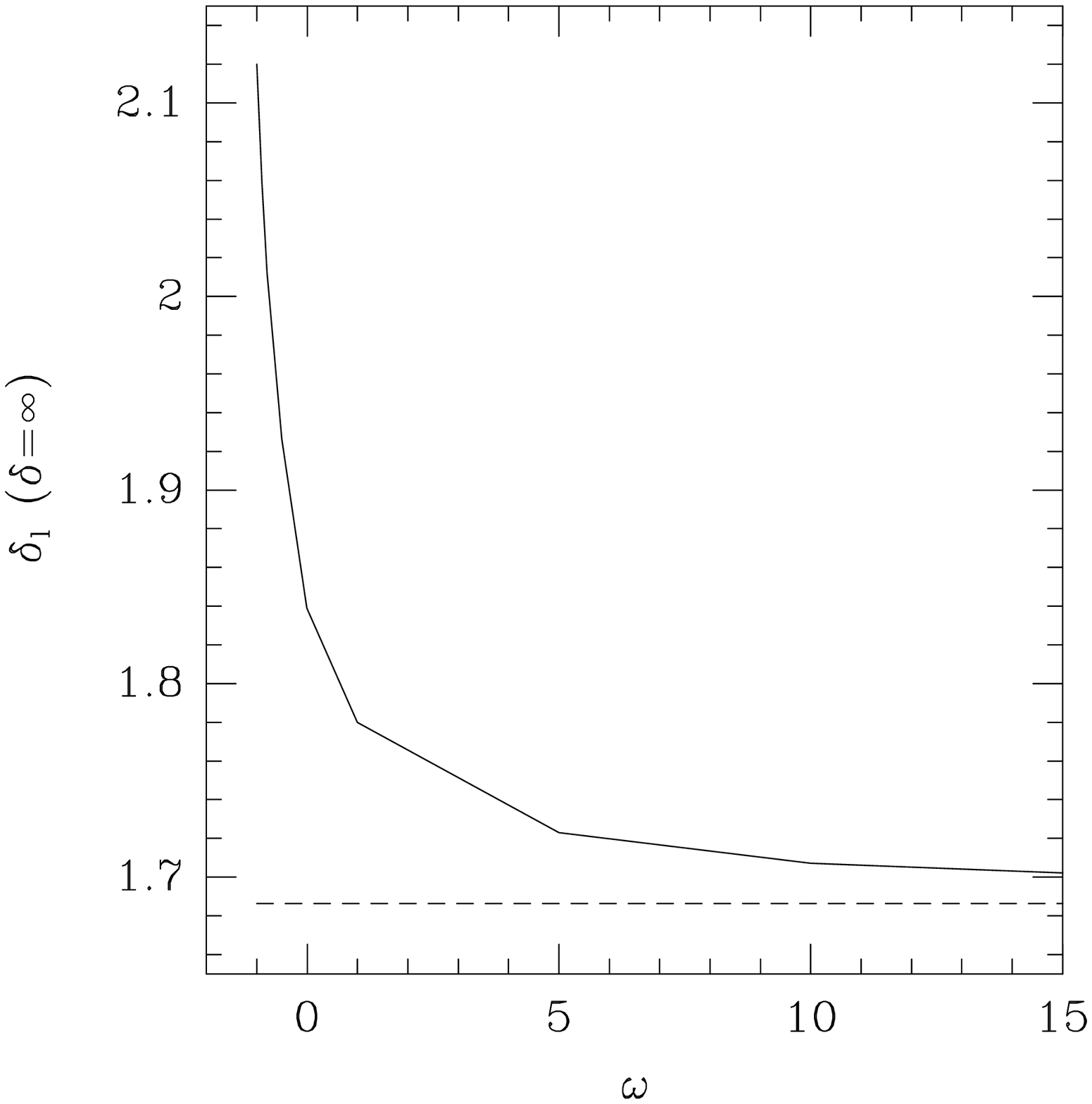}}
\figcaption{Left panel: dashed lines show $\nu_2$ (top)
and  $\nu_3$ (bottom) as a function of $\omega$
for a time varying gravitational constant 
$G = G_0~ a^{-1/(1+\omega)}$. The GR results, $G=G_0$,
(continuous horizontal lines) are bracket by 
$10\%$ errors (dotted lines).
Right panel: the critical value of the linear density contrast
$\delta_c$ where $\delta = \infty$ as a function
of $\omega$.
\label{nu23}}
\end{figure*}


In the non-linear case we consider the full version of equation
[\ref{33.7}].
We can now proceed as before, using the expansion in Eq.[\ref{loclag}]
with $\delta_l$ given by the linear growth factor
$D=a^{\alpha_1}=e^{\alpha_1\eta}$, and compare order by order. We find

\bea
\nu_2 &=& {{34\omega + 56}\over{21 \omega + 36}}
~= ~ {34\over{21}} ~\left[1 - {8\over{119}} ~{1\over{\omega}} 
+ \Or\left({1\over{\omega^2}}\right)\right]
\label{nu2w} \\[1 ex]
\nu_3  &=& {2(944 + 1136 ~\omega + 341 ~\omega^2)
\over{3(12+7\omega)(16+9\omega)}} \label{nu3w} \nn \\
~ &=& ~ {682\over{189}} ~\left[1+{3452\over{21483}} ~{1\over{\omega}} 
+ \Or\left({1\over{\omega^2}}\right)\right] 
%
\eea

Note how for positive $\omega$ non-linear effects tend to
compensate the increase in linear effects, cf Figure \ref{alphas},
whereas for $\omega<-4/3$, the linear effects are reduced
($\alpha<1$) while non-linearities get larger.

Figure~\ref{nu23} shows the variation in $\nu_2$
 as a function of $\omega$ using Eq.[\ref{nu2w}]. 
Negative values of $\omega$ produce almost symmetrical variations in
the opposite direction when $|\omega|$ is large. For small $\omega$
there is a pole at $\omega=-12/7$
where $\nu_2$ diverges. But note that there is no
growing linear mode in this case, which means that
fluctuations are rapidly suppressed.

\subsubsection{Strongly non-linear regime} 
\label{sec:deltac}

Figure \ref{deldell} shows the fully non-linear solution
for the overdensity  $\delta$ as a function of the 
linear one $\delta_l$. The continuous line shows the
standard solution to Eq.[\ref{14}] as given 
in Eq.[\ref{SC1}]-[\ref{SC2}]. 
As can be seen in the
Figure, there is a critical value of
$\delta_l=3/2(3\pi/2)^{3/2} \simeq 1.6865$ where the
non-linear fluctuations become infinite. This corresponds
to the point where the spherical collapse occurs (see Peebles 1993).
Thus an initial fluctuations $\delta_0$ will collapse after evolving
a time $t$, such that the growth factor is 
$D(t) = \delta_c / \delta_0$. For the standard GR, flat and matter
dominated case, this time would correspond to a formation red-shift:
$z_f =  \delta_0 / \delta_c -1$
(if we use $a=1$ today). For the BD case both $\delta_c$ and
$D(t)$ are different, so that formation times $z_f$ will be 
correspondingly different (see Eq.[\ref{zf}]).
The short-dashed lines in Figure \ref{deldell}
correspond to the same exact solution in the
BD model with $\omega=10$ and $\omega=1$. 
Right panel in Figure \ref{nu23} illustrates how $\delta_c$ changes in the
BD model as a function of $\omega$.


\subsection{Gravitational Growth with $H^2 \sim a^{-3(1+\epsilon)}$}
\label{sec32}

Consider now the flat case with $\Omega_k=\Omega_\Lambda=0$.
To account for a simple variation on the standard Einstein's field
equations we will consider the case where fluctuations
grow according to the matter dominated case
(ie $\gamma=0$) but the background evolves in a different way.
We will assume that the Hubble rate goes like
$H^2 \sim a^{-3(1+\epsilon)}$ rather than $H^2 \sim a^{-3}$.
It might be possible to find some motivation for this model,
but this is beyond the scope of this work. Here we just want to
introduce some parametric variations around the standard field
equations  to see how things might
change. In this case we have:

\beq
{d^2 \delta \over{d^2\eta}} + {1-3\epsilon \over{2}}~ {d\delta\over{d\eta}}
- {3\over{2}}  ~\delta = {4\over{3}} {1\over 1+\delta}
\left({d \delta \over{d\eta}}\right)^2
~+~ {3\over{2}} ~\delta^2 
\eeq

The solutions for the linear growth factor index and the non-linear
coefficient $\nu_2$ are

\bea
\alpha_1 &=& \frac{-1+3\epsilon + \sqrt{25-6\epsilon+9\epsilon^2}}4
\label{eqalphae}        \\
\nu_2    &=& \frac{131-30\epsilon+45\epsilon^2 + (1-3\epsilon)\,
\sqrt{25-6\epsilon+9\epsilon^2}}{84 -18\epsilon+27\epsilon^2}
\label{nu2e}
\eea

\begin{figure*}
\centerline{\epsfxsize=8truecm
\epsfbox{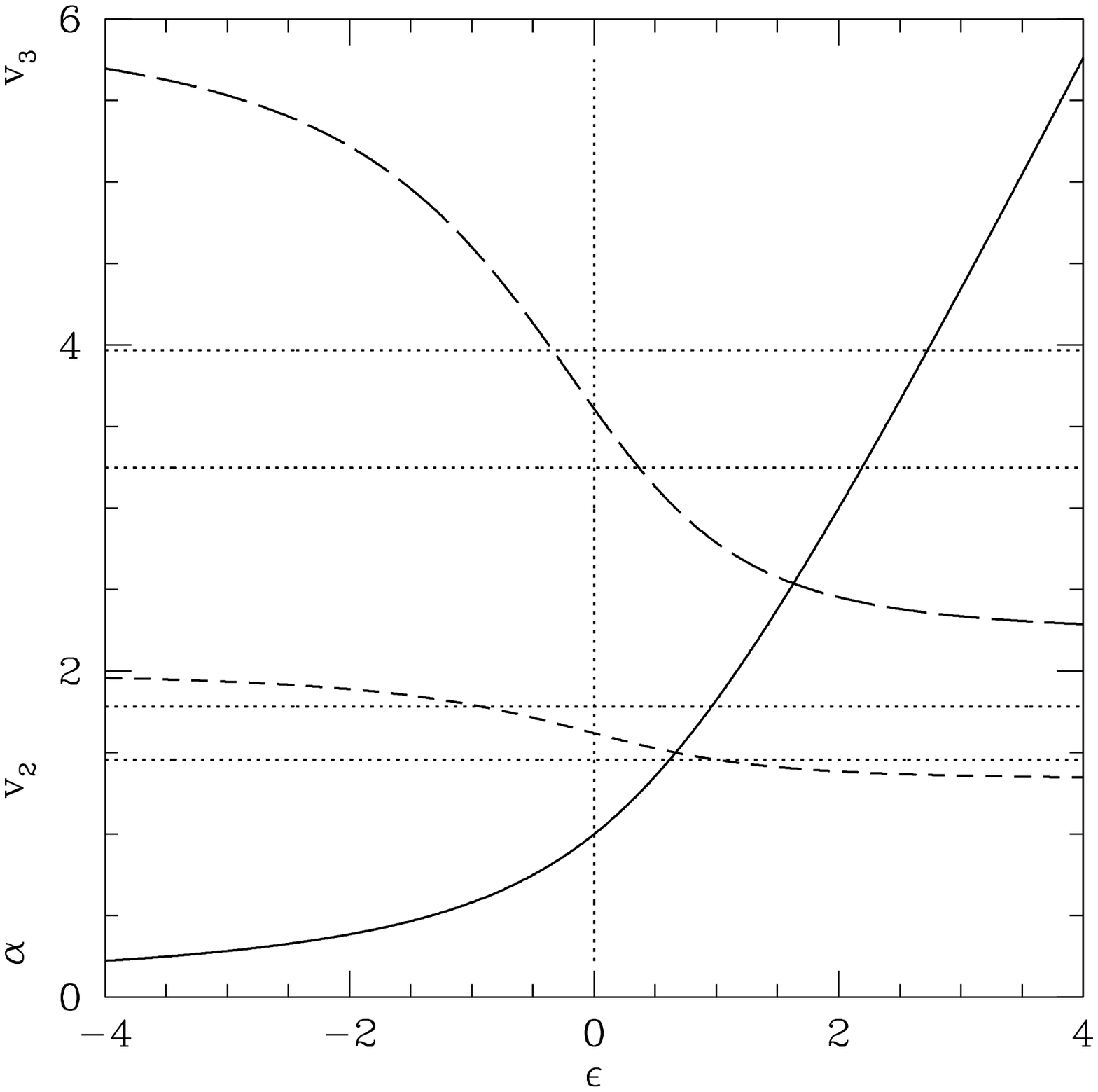}\epsfxsize=8truecm
\epsfbox{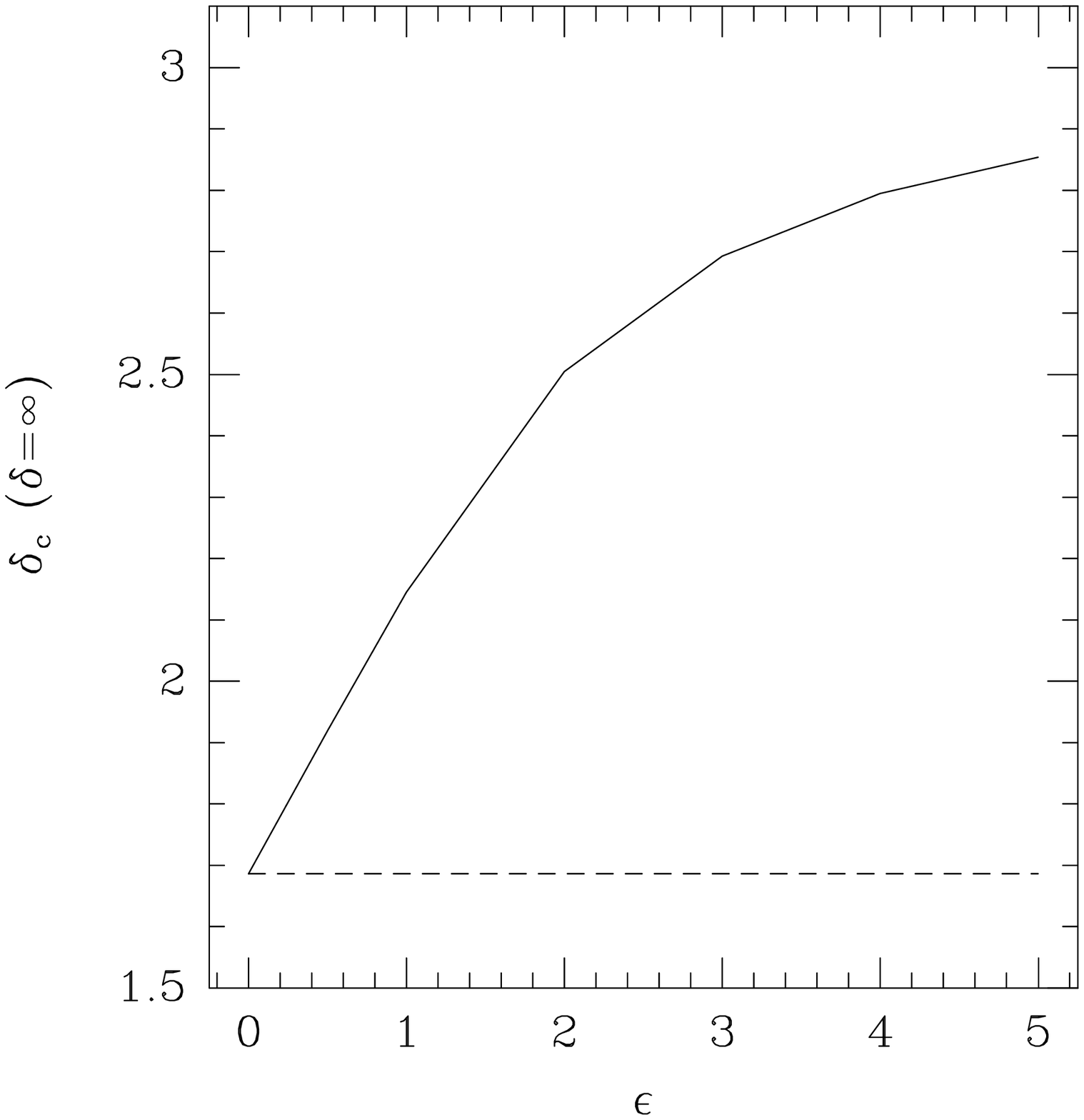}}
\figcaption{Left panel: 
The linear growth index $\alpha_1$ (continuous line) and
non-linear coefficients $\nu_2$ (short-dashed) and $\nu_3$ (long-dashed),
as a function of $\epsilon$, which parameterizes a non-standard Hubble rate
$H^2 \sim a^{-3(1+\epsilon)}$. Vertical dotted line corresponds to the
standard Hubble law ($\epsilon=0$). The horizontal dotted lines bracket
the $\nu_2$ and  $\nu_3$ regions within $10\%$ error of the standard
($\epsilon=0$) case. 
Right panel: the critical value of the linear density contrast
 $\delta_c$ where $\delta = \infty$ as a function
of  $\epsilon$.
\label{alphae}}
\end{figure*}

These solutions as a function of $\epsilon$ are illustrated in
Figure~\ref{alphae}, which also shows $\nu_3$. As can be seen
in the Figure, the higher the linear growth index $\alpha_1$
the lower the non-linear coefficients. 

Right panel in Figure \ref{alphae} shows the corresponding variation
in $\delta_c$.

\section{Observational consequences}

We will focus here on Gaussian initial conditions. That is, our initial field
for structure formation is a spatial realization of a (three-dimensional)
Gaussian distribution with a given power spectrum shape, and a very small
initial amplitude. As we are interested in the gravitational regime alone,
this field will be smoothed over a large enough scale, corresponding to the
distance beyond which non-gravitational forces (eg hydrodynamics) can be
neglected. Thus at each point the overdensity $\delta(\x)$ grows according 
to gravity, which in the shear free approximation is just a local dynamics:
the spherical collapse (eg Eq.[\ref{14}]).

\subsection{Cumulants}
\label{sec:cumulants}

Consider  the $J-order$ moments of the fluctuating field:

\beq
m_J \equiv \lexp \delta^J \rexp .
\eeq

Here the expectation values $\lexp ... \rexp$ correspond to an average over
realizations of the initial field. On comparing with observations we assume
the {\it fair sample hypothesis} (\S 30 Peebles 1980), by which we can
commute spatial integrals with expectation values. Thus, in practice
$\lexp ... \rexp$ is the average  over positions in the survey area. In
this notation the variance is defined as:

\beq
Var(\delta) \equiv \sigma^2 \equiv m_2 - m_1^2
\eeq

More generally, we introduce the {\em connected moments} $\xibar_J$, which
carry statistical information independent of the lower order moments, and
are formally denoted by a bracket with subscript $c$:

\beq
\xibar_J \equiv \lexp \delta^J \rexpc
\eeq

The connected moments are also called {\em cumulants}, {\em reduced} moments
or {\em irreducible} moments. They are defined by just subtracting the lower
order contributions:

\bea
\xibar_1 &=& m_1 \equiv  0 \nonumber \\*
\xibar_2 &=& \sigma^2 = m_2 - \xibar_1^2 = m_2 \nonumber \\*
\xibar_3 &=& m_3 -3 \xibar_2 \xibar_1 - \xibar_1^3 = m_3  \\*
\xibar_4 &=& m_4-4 \xibar_3 \xibar_1-3 \xibar_2^2 -6 \xibar_2
\xibar_1^2-\xibar_1^4 = m_4 -3 m_2^2 \nonumber 
\label{connect}
\eea
and so on. It is useful to introduce the {\em hierarchical ratios}:

\beq
S_{J}\,=\,\frac{\xibar_J}{{\xibar_2^{J-1}}}
\label{sj}
\eeq
which  are also called normalized one-point cumulants or reduced cumulants.
We shall use the term {\it skewness}, for $S_3= \xibar_3 / \xibar_2^2$
and {\it kurtosis}, for $S_4= \xibar_4 / \xibar_2^3 $.

\subsubsection{Linear Theory}

As mentioned in \S\ref{sec:linear}, initial fluctuations, $\delta_0$, no
matter of what amplitude, grow all by the same factor, $D$; thus the
statistical properties of the initial field are just linearly scaled in
the final (linear) field, $\delta_l$:

\beq
\lexp \delta^J_l \rexpc = D^J ~ \lexp \delta_0^J \rexpc
\eeq

Consider for example the linear rms fluctuations $\sigma_l$ or its variance
$\sigma^2_l$. In the linear regime we have:

\beq
\sigma^2_l ~\equiv~ \lexp \delta^2(t)\rexp
= \lexp D(t-t_0)^2 \delta_0^2 \rexp = D(t-t_0)^2 ~\sigma_0^2
\eeq
where $\sigma_0$ refers to some initial reference time $t_0$. To give an
idea of this effect, consider the growth of fluctuations since matter
domination, when the universe was about 1100 times smaller. In General
Relativity (GR) in the matter dominated Einstein-deSitter universe,
$\sigma$ would grow by a factor $D\simeq 1100$. While, if we take
$\omega\simeq 10$ in the DB theory, eg Eq.[\ref{eq:alphaw}], we have that
fluctuations increase instead by a factor $D\simeq 2079$, which is about
$1.9$ times larger in $\sigma$, so the variance nowadays would be about
$3.6$ times larger if we fixed it around the COBE variance of Cosmic
Microwave Background (CBM) temperature fluctuations. For $\omega\simeq 100$,
the variance would only be $14\%$ larger than in GR. This latter result is
small, but it could be relevant for future precision measurements (eg MAP
or PLANCK satellites to map CMB and 2DF or SLOAN DIGITAL SKY galaxy surveys).
Similar considerations can be made for the values of $\alpha$ with a
different cosmic equation of state, eg Eq.[\ref{alpha1g}] or a different
Hubble law, Eq.[\ref{eqalphae}]. In general we can write that a small change
in $\alpha$ would produce a relative change in the linear rms of

\beq
{\Delta \sigma \over{\sigma}} = \ln(1+z) \Delta\alpha
\eeq

Thus, a change of only $1\%$ in the absolute value of the equation of state
$\gamma$, would produce a relative change of $20\%$ in $\sigma$ between
recombination $z\simeq 1100$ and now, cf Eq.[\ref{alpha1g}].

The hierarchical ratios (see Eq.[\ref{sj}]) will scale as,
$S_J = {S_J(0)/{D^{J-2}}}$, where $S_J(0)$ are the initial ratios. This
implies that the linear growth erases the initial skewness and kurtosis,
so that $S_J \rightarrow 0$, as time evolves (and $D \rightarrow \infty$).
Note that if we want to do a meaningful calculation of these ratios or the
cumulants, in general we might need to consider more terms in the
perturbative series, Eq.[\ref{loclag}]. For Gaussian initial conditions
$S_J(0) = 0$, and we need to consider higher order terms in the perturbation
series to find the leading order prediction.

\subsubsection{Weakly non-linear}
\label{sec:gausstree}

The next to leading order solutions for the cumulants of the evolved
field given the expansion Eq.[\ref{loclag}], can be easily found by just
taking expectation values of different powers of $\delta$ (see eg Fosalba
\& Gazta\~naga 1998a). For leading order Gaussian initial conditions we have

\bea
S_{3} &=& 3 \nu_2 + \Or(\sigma_l^2) \nn \\
S_{4} &=& 4 \nu_3 + 12 \nu_2^2 + \Or(\sigma_l^2)
\label{s3nu2}
\eea

For non-Gaussian initial conditions see Fry \& Scherrer (1994) Chodorowski
\& Bouchet (1996), Gazta\~naga \& Mahonen (1996), Gazta\~naga \& Fosalba
(1998).

If we use for $\nu_2$ the solution in Eq.[\ref{nusc}], eg $\nu_2=34/21$,
the skewness yields  $S_3= 3 \nu_2= 34/7$, which reproduces the exact 
perturbation theory (PT) result by Peebles (1980) in the matter dominated
Einstein-deSitter universe. Thus the shear-free or SC model gives the exact
leading order result for the skewness. This is also true for higher orders
(see Bernardeau 1992 and Fosalba \& Gazta\~naga 1998a) and for other
cosmologies (eg Bouchet et al. 1992,  Bernardeau 1994a, Fosalba \&
Gazta\~naga 1998b, Kamionkowski \& Buchalter 1999). For smoothed fields,
the exact leading order results are slightly different:

\bea
S_3 &=& {34\over 7} + \gamma_1 \\
S_4 &=&  {60712\over 1323} + {62\over 3}\,\gamma_1 + 
{7\over 3}\,\gamma_1^2
\label{PTpred}
\eea
where $\gamma_1$ is the logarithmic slope of the smoothed variance
(see Juszkiewicz \etal 1993, Bernardeau 1994a, 1994b). These can also
be reproduced in the shear-free approximation as shown by Gazta\~naga
\& Fosalba (1998); this results in a smoothing correction:

\bea
\overline{\nu_2} &=& \nu_2 + {\gamma_1 \over 3} \nn \\
\overline{\nu_3} &=&{1 \over 4}(-2 \,\gamma_1 + \gamma_1^2 +
6 \,\gamma_1\, \nu_2 + 4\,\nu_3)
\eea  
and replacing $\nu_2$ and $\nu_3$ by $\overline{\nu_2}$ and
$\overline{\nu_3}$ in Eq.[\ref{s3nu2}] (see Fosalba \& Gazta\~naga 1998a
for more details). There are also corrections to the above expressions
when measurements are taken in red-shift space (eg Hivon et al 1995,
Scoccimarro, Couchman and Frieman 1999). Next to leading order terms have
been estimated by Scoccimarro \& Frieman (1996) (see also Fosalba \&
Gazta\~naga 1998a,b).

The smoothed values of $S_3$ and $S_4$ can be measured as traced by the
large scale galaxy distribution (eg Bouchet \etal 1993, Gazta\~naga 1992,
1994, Szapudi el at 1995, Hui \&  Gazta\~naga 1999 and references therein),
weak-lensing (Bernardeau, Van Waerbeke \& Mellier 1997, Gazta\~naga \&
Bernardeau 1998, Hui 1999) or the Ly-alpha QSO absorptions (Gazta\~naga \&
Croft 1999). These measurements of the skewness and kurtosis can be
translated into estimations of $\nu_2$ and $\nu_3$ which can be used to
place constraints on $\gamma$, $\omega$ or $\epsilon$ using Eq.[\ref{nu2g}],
[\ref{nu2w}] and [\ref{nu2e}]. For small values of these parameters the
relationship is linear, so the uncertainties in $S_3$ and $S_4$ would
directly translate into the corresponding  uncertainties in $\gamma$,
$\omega$ or $\epsilon$. 

The expressions above apply to unbiased tracers of the density field; since
galaxies of different morphologies are known to have different clustering
properties, at least some galaxy species must be biased tracers of the mass.
As an example, suppose the probability of forming a luminous galaxy depends
only on the underlying mean density field in its immediate vicinity. Under
this simplifying assumption, the relation between the galaxy density field
$\delta_{gal}(\x)$ and the mass density field $\delta(\x)$ can be written as

\beq
\delta_{gal}(\x) = f(\delta(\x)) = \sum_n ~{b_n \over {n!}} ~~\delta^n(\x),
  \nn
\eeq
where $b_n$ are the bias parameters. Thus, biasing and gravity could produce
comparable non-linear effects. To leading order in $\xibar_2$, this local
bias  scheme implies $\xibar_2^{gal} =b_1^2 \xibar_2$, and (see Fry \&
Gazta\~naga 1993)

\bea
S_3^{gal} &=& {S_3 \over{b_1}} + 3~ {b_2 \over{b^2_1}} \nn \\
S_4^{gal} &=& {S_4 \over{b_1^2}} + 12 {b_2 S_3\over b_1^3} +
4 {b_3\over b_1^4} + 12 {b_2^2\over b_1^4} 
\eea

Gazta\~{n}aga \& Frieman (1994) have used the comparison of $S_3$ and
$S_4$ in PT with the corresponding values measured APM Galaxy Survey
(Maddox et al. 1990), to infer that $b_1 \simeq 1$, $b_2 \simeq 0$ and
$b_3 \simeq 0$, but the results are degenerate due to the relative
scale-independence of $S_N$ and the increasing number of biasing
parameters. One could break this degeneracy by using the configuration
dependence of the projected 3-point function, $q_3(\alpha)$, as proposed
by  Frieman \& Gazta\~naga (1994), Fry (1994),  Matarrese, Verde \&
Heavens (1997), Scoccimarro \etal (1998).
As shown in Frieman \& Gazta\~naga (1999), the configuration dependence
of $q_3({\alpha})$ on large scales in the APM catalog  is quite close to
that expected in perturbation theory (see Fry 1984, Scoccimarro \etal 1998,
Buchalter, Jaffe \& Kamionkowski 2000),
suggesting again that $b_1$ is of order unity (and $b_2 \simeq 0$) for these
galaxies. These agreement  indicates that large-scale structure is driven by
non-linear gravitational instability and that APM galaxies are relatively
unbiased tracers of the mass on these large scales.

The values of $S_3$ and $S_4$ in the APM are measured to agree with the
standard matter dominated Einstein-deSitter universe within about
$10\%-20\%$ (see Gazta\~naga 1994; Gazta\~naga \& Frieman 1994; Baugh,
Gazta\~naga \& Efstathiou 1995; Gazta\~naga 1995, Hui \& Gazta\~naga 1999),
also in agreement with the shape information in the 3-point function (see
Frieman \& Gazta\~naga 1999). For example, using the projected APM catalogue
Gazta\~naga 1994 (Table 3) finds an average of $S_3 = 3.2 \pm 0.2$
and $S_4 \pm 20.6 \pm 2.6 $ scales between 7 and 30 $\Mpc$.
For an average APM slope of $\gamma_1 \simeq 1.7$, these values are
in agreement with the PT predictions in
Eq.[\ref{PTpred}] yield: $S_3 \simeq 3.1$ and $S_4 \simeq 18$.

The 1--sigma error bar of $\simeq 10\%$ on large
scales quoted by Gazta\~naga 1994 is mostly statistical (sampling error).
Other systematics effects due to biasing, projection, or large scale errors
in the building of the APM catalogue could be of the same order (see Frieman
\& Gazta\~naga 1999 and Hui \& Gazta\~naga 1999). Thus given the current
uncertainties it would be conservative to take a $20\%$ error bar.
Unfortunately, with such large error bars we  can not constraint much the
values of $\gamma$  $\omega$ or $\epsilon$. Stronger constraints can be
found if we take the more optimistic 1--sigma $10\%$ error bars in the
measurements of $S_3$ and $S_4$. This case is shown as horizontal dotted
lines in Figures \ref{ggamman}, \ref{nu23} and \ref{alphae}. From $\nu_2$
the $10\%$ uncertainty translates into

\bea
-0.2 < &\gamma & < 0.4 \nn \\
-2.4 > &\omega & > -1.0 \nn \\
-0.9 < &\epsilon & < 0.9 
\label{bounds1}
\eea

Note that this is still of marginal interest. For example, the constraints
on $\gamma$ include the possibility of a radiation ($\gamma=1/3$), matter
($\gamma=0$) or negative pressure $\gamma<0$. From $\nu_3$ we can obtain
stronger constraints from a $10\%$ error (but obviously systematic effects
could be larger for higher order cumulants):

\bea
-0.1 < &\gamma & < 0.15         \nn \\
-3.4 > &\omega & > -0.2         \nn \\
-0.35 < &\epsilon & < 0.35      \label{bounds2}
\eea

These bounds are more interesting. It is clear that forthcoming surveys
(such as the SLOAN Digital Sky Survey) will dramatically improve this
situation (for errors on statistics  see Szapudi, Colombi and Bernardeau
1999, and references therein).

Note that the above results are independent of the normalization of
fluctuations.

\subsection{Collapsed objects}
\label{sec:abundances}

\begin{figure*}
\centerline{\epsfxsize=8truecm
\epsfbox{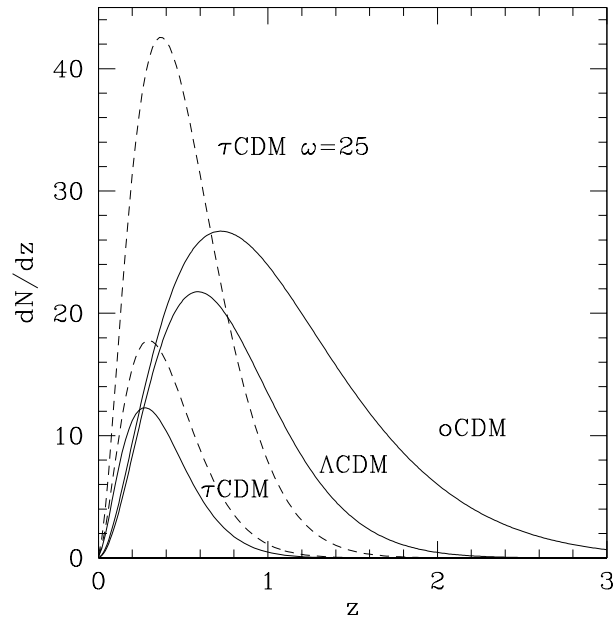}}
\figcaption[junk]{The solid lines represent the expected differential count
distribution per square degree of massive clusters ($\rm M > M_{th}
\times 10^{14}\,h^{-1}_{50}{M_{\sun}} $) for three cosmologies oCDM
($M_{th}=1.9 $, $\Omega_{m}=0.3,\Omega_{\Lambda}=0, h=0.65,
\Gamma=0.25, \sigma_{8}=1.0$); $\Lambda$CDM ($M_{th}=2.2$,
$\Omega_{m}=0.3,\Omega_{\Lambda}=0.7, h=0.65, \Gamma=0.25,
\sigma_{8}=1.0$) and $\tau$CDM ($M_{th}=1.3$,
$\Omega_{m}=1.0;\Omega_{\Lambda}=0, h=0.5, \Gamma=0.25,
\sigma_{8}=0.56$) derived using the Press-Schechter prescription.
The lower (upper) dashed lines correspond to the
a Brans-Dicke Cosmology with $\omega=100$ ($\omega=25$) normalized to
COBE with the $\tau$CDM model.
\label{clustera}.}
\end{figure*}

Press \& Schechter (1974) formalism and its extensions (eg Bond \etal 1991;
Lacey \& Cole 1993) predict the evolution of the mass function of halos and
also their clustering properties. Comparison with N-body simulations show a
very good agreement of these prescriptions for a wide range of statistical
properties (eg see Lacey \& Cole 1994 and references therein). For example,
the comoving number density of collapsed objects (halos or clusters) of mass
$M$ is

\begin{equation}
n(M) dM = - \sqrt{2\over \pi} 
\, \left({\delta_c \over \sigma}\right) {d \ln \sigma  \over 
d \ln M} \, \exp{\left(-{\delta_c^2 \over 2 \sigma^2}\right)} \, 
{\bar\rho \, dM \over M^2}
\label{PS}
\end{equation}
where $\sigma=\sigma(R)$ is the current linear rms fluctuation at the scale
$R$ corresponding to the mass $M= 4/3\pi R^3 \bar\rho$, and $\bar\rho$ is
the mean background. The value of $\delta_c$ corresponds to the value of
the linear overdensity at the time of collapse. The collapsing structure
virializes when the (non-linear) overdensity becomes very large
($\delta \ga 100$). The actual definition is not very important, as once
$\delta \ga 100$, the non-linear collapse is quite rapid, as can be seen
in the plots of Figure \ref{deldell}, and the corresponding value of
$\delta_l$ does not change much. Here we will take $\delta_c$ to be the
critical value where $\delta \rightarrow \infty$; other prescriptions (eg
the value of $\delta_l$ corresponding $\delta\simeq 178$) yield similar
results. For the standard Einstein-de Sitter case we have
$\delta_c \simeq 1.686$. Note that the above abundance depends on the ratio

\beq
\nu \equiv {\delta_c \over{\sigma}}
\label{nu}
\eeq

The time of collapse or formation is just given by the ratio of $\delta_c$
to the linear overdensity $\delta_l$ today

\beq
z_f = \left({\delta_l\over{\delta_c}}\right)^{1/\alpha} -1
\label{zf}
\eeq
so that an object which has $\delta_l=\delta_c$ now, has a formation red-shift
$z_f=0$, while a fluctuation 4 times larger collapses at $z_f=3$ if $\alpha=1$
or at $z_f=1$ if $\alpha=2$.

Non-standard parametrisation of the spherical collapse considered in the
previous sections can change the above formalism in two ways. If we label
objects by its {\it initial} overdensity $\delta_0$ then the corresponding
$\delta_l$ today is

\beq
\delta_l = \delta_0\,a^\alpha
\eeq

So a different value of $\alpha$, from the standard GR result
($\Delta\alpha \equiv \alpha-\alpha_{GR}$), as shown in Figures~\ref{ggamman},
\ref{alphas}, and~\ref{alphae}, will produce a different amplitude of linear
fluctuations today. Moreover, as shown in \S\ref{sec:deltac} and
Figures~\ref{deltacg}, \ref{nu23}, and~\ref{alphae}, the solution to the
spherical collapse equation produces different values of $\delta_c$, and
therefore different mass functions and formation times.
Finally, for a directly measurable quantity, such as the surface
density of objects, typically one needs the volume element, which is also
a function of the cosmology.

For example if fluctuations are normalized at a given red-shift, $z_n$,
then the change in $\delta_l$ today will be

\beq
{\Delta \delta_l \over{\delta_l}} = \Delta\alpha \, \log{(1+z_n)}
\eeq

For  recombination, eg COBE normalization, we have $z_n \simeq 1100$, and

\beq
{\Delta \delta_l \over{\delta_l}} \simeq  3 \, \Delta\alpha
\eeq

In the case of the BD theory, we can see in Figure \ref{alphas} that for
$\omega>-1$,  $\Delta \alpha \equiv \alpha-\alpha_{GR}>0$, which means that
$\Delta \delta_l>0$. This makes sense as the linear growth is faster and,
for fixed initial fluctuations, the final linear overdensity will be larger.
As shown in right panel of 
Figure \ref{nu23}, $\delta_c$ will also be larger. Thus in
this case the effects tend to compensate each other. This is true for both
the formation red-shift $z_f$ or for $\nu$ in Eq.[\ref{nu}] above. For the
formation red-shift $z_f$ we have

\beq
{\Delta z_f \over{1+z_f}} \simeq {1\over{\alpha}} \left(
{\Delta \delta_l \over{\delta_l}}
- {\Delta \delta_c \over{\delta_c}} \right)
\eeq
which is only valid for small changes. In the BD example given above with
$\omega=10$ (and COBE normalization) we have that
${\Delta \delta_l /{\delta_l}} \simeq 0.9$ while 
${\Delta \delta_c /{\delta_c}} \simeq 0.01$, so the net effect is
still quite large. In this case, a formation red-shift of $z_f=1$
will change to  $z_f=1.39$. Thus, a positive finite $\omega$ 
(which corresponds to a larger $G$ at high red-shifts) tends to produce
larger (earlier) formation red-shifts and higher densities (or larger
abundances) at a given red-shift, than the standard model. This goes in
the direction of some recent observations (eg see Bahcall \& Fan 1998;
Robinson, Gawiser \& Silk 1998, Willick 1999), which seem to need larger
abundances that expected in some standard cosmologies. This interpretation
is degenerate with respect to initial conditions and cosmological parameters.

Figure \ref{clustera} illustrates the large
differences in the cluster counts that
can be seen between different cosmological models at $z>1$
(see Holder et al.~1999 for details).
Deviations from General Relativity in the BD models with
$\omega=100$ and $\omega=25$ can be noticed even 
at low redshift, when models are normalized to CMB fluctuations.

A similar trend is found for the case of Hubble rate
$H^2= a^{-3(1+\epsilon)}$ parametrisation. A change of
$|\Delta \epsilon| \simeq 0.3$ (allowed by the bounds
in Eq.[\ref{bounds2}]), when normalized to COBE, also
produces $|{\Delta \delta_l /{\delta_l}}| \simeq 0.9$
and a smaller effect on ${\Delta \delta_c /{\delta_c}}$. This
translates into a similar change (of several tens to hundreds of
percent) in $z_f$. Earlier (later) formation times and larger
(smaller) abundances are found for $\epsilon>0$ ($\epsilon<0$).

The change in the equation of state $p=\gamma \rho$ could produce
comparable effects. The allowed values in Eq.[\ref{bounds2}] of
$|\Delta \gamma| \simeq 0.1$ translate into
$|{\Delta \delta_l /{\delta_l}}| \simeq 0.3$, which results in similar
changes for $z_f$ in either direction, with earlier formation for $\gamma>0$.

If the normalization is not fixed, ie we do not quite know what is
the value of the initial fluctuation that gave rise to an object
we see today (eg a cluster), then all the relative change in
the formation or abundance comes through $\delta_c$, which tends to
produce smaller (later) formation red-shifts ($\delta_c$ is larger
than the standard GR value) and lower densities (or smaller abundances)
at a given red-shift. 

\section{Discussion and Conclusions}

We have reconsidered the problem of non-linear structure formation in
two different contexts that relate to observations: 1-point cumulants
of large scale density fluctuations and the epoch of formation and
abundance of structures using the Press \& Schechter (1974)
formalism. We have use the the shear-free or spherical collapse (SC)
model, which is very good approximation for the above applications.
We have addressed the question of how different are the predictions
when using a non-standard theory of Gravity, such as BD model, or
non-standard cosmological model (eg a different equation of state or
Hubble law). Note that these are slight variations on the standard
theme in the sense that they preserved the main ingredients of GR,
such as the covariance and the geometrical aspects of the theory,
including the same metric, with only slight changes in the field
equations.

We have also presented some preliminary bounds on $\gamma$, $\omega$ and
$\epsilon$ from observations of the skewness and kurtosis in the APM Galaxy
Survey, eg Eq.[\ref{bounds1}]-[\ref{bounds2}]. These bounds are optimistic
given the current data, but the situation is going to  change rapidly,
and one can hope to find much better bounds form upcoming data
(such as 2DF or SDSS projects). In terms of the equation of state
the bounds in Eq.[\ref{bounds2}] would indicate that our Universe
is neither radiation ($\gamma=1/3$) or vacuum dominated ($\gamma=-1$),
but somewhere in between (eg matter dominated). In terms of the
Gravitational constant, the bounds on $\omega$  from  Eq.[\ref{bounds2}]
would say that $G$ has not changed by more than $\simeq 5\%$ from
$z \simeq 1.15$, or by distances of $\simeq 400 \Mpc$. Clustering at
higher red-shift would probe much larger scales and times. In terms of
$\epsilon$ the bounds Eq.[\ref{bounds2}], would say that the  Hubble law
does not differ by more than $7\%$ from the standard result (assumed here
to be $\epsilon=0$). We have also shown how halo and cluster abundances
and formation times could change in these non-standard cases. The above
bounds on $\gamma$, $\omega$ and $\epsilon$ from observations of the
skewness and kurtosis in the APM still allow significant changes (of
several tens to hundreds of percent) on formation red-shifts $z_f$ and
the corresponding abundances (see \S\ref{sec:abundances}).

In the context of BD models the limits we find for $\omega$ are less
restrictive than the solar system limits $\omega \ga 100$. However, BD
models allow $\omega=\omega(\phi)$ so that $\omega$ can increase with
cosmic time, $\omega=\omega(z)$, in such a way that it could approach
the general relativity predictions ($\omega \rightarrow \infty$) at
present time and still give significant deviations at earlier
cosmological times. It is important to recall that our theory of
gravity has only be tested on stellar distances (a.u.) while we want
to use it on cosmological scales ($Mpc$). Our working example
shows, for the first time, how non-linear effects are changed in such
a model and sets the framework to study non-linear effects of more
complicated (or realistic) Scalar-Tensor theories of gravity.

It is straightforward to combine several of the changes proposed
here to explore more general situations. One could for example
parameterize theories in the ($\gamma, \omega$) plane, eg different
equations of state with different BD parameters, or consider
the whole ($\gamma,\omega,\Omega_M,\Omega_\Lambda$) space.
One could also consider 
a different equation of state for the  $\Lambda-$component, 
as in quintessence cosmologies (Caldwell, Dave, Steinhardt 1998), 
such models  have already been used to predict cluster abundances
within the ``standard'' cosmology (see Haiman, Mohr, Holder 2000 and
references therein).
This would obviously allow for a wider set of possible solutions
and degeneracies. One should also consider other observational
consequences of these variations, in particular relating to BD
theory, such as the age of the Universe, the effects on CMB (eg see
Chen \& Kamionkowski 1999), radiation-matter transition (Liddle,
Mazumdar \& Barrow 1998), or the constraints from nucleosynthesis
(Santiago \etal 1997). These considerations could rule out some aspects
of the proposed variations on the standard model, or might require more
elaborate solutions (eg $\omega=\omega(\phi)$ which implies
$\omega =\omega(z)$). But even if this were the case, we still have
learn a few new things about how structure formation depends on the
underlying theory of Gravity, which is a first step towards further
analysis of these issues.

Throughout this paper we have assumed Gaussian initial conditions
and no biasing. Both biasing (eg Fry \& Gazta\~naga 1993) and
non-Gaussianities in the initial conditions (Gazta\~naga \& Fosalba 1998)
would  provide an additional source of degeneracy as they might produce
similar effects as the non-standard variations presented here. This is
the case for example when we have non-zero initial skewness or kurtosis,
which could produced quite different values of $S_3$ and $S_4$
(eg see Gazta\~naga \& Mahonen 1996; Peebles 1999a,b; White 1999;
Scoccimarro 2000), and therefore to the inferred values of $\nu_2$ and
$\nu_3$. Biasing can have a very similar effect (eg see Mo, Jing \& White
1997). One would also expect some level of degeneracy with biasing and
initial conditions for cluster abundances or formation times 
(see Robinson, Gawiser \& Silk 1998, Willick 1999).

Rather than proposing an alternative theory of gravity or cosmological
model, the aim of this paper was to show that some small deviations
from the current paradigm have significant and measurable consequences
for non-linear structure formation. This could eventually help
explaining some of the current puzzles confronting the theory, such as
the need of non-baryonic dark matter. Alternatively, current and
upcoming observations of non-linear clustering and mass functions can
be used to explore our assumptions and place limits on the theory of
gravity at large ($\ga 1 \Mpc$) scales. This provides an interesting
test for gravity as the driving force for structure formation and for
our knowledge of the cosmological equation of state.  A more
comprehensive comparison with particular scenarios is left for future
work.

\section*{Acknowledgments}

One of us (JAL) gratefully acknowledges financial support from the Spanish
Ministry of Education, contract PB96-0384, and also Institut d'Estudis
Catalans.  EG acknowledges support from  CSIC, DGICYT (Spain), project
PB96-0925. We would like to thank IEEC, where most of this work was carried
out. 


\end{document}